\documentclass[10pt,aps,amsmath,showkeys]{revtex4}

\usepackage{xcolor}
\usepackage{latexsym,amsmath,amssymb,amsbsy,graphicx}

\makeatother

\begin{document}
\title{Multiple-Timescale Theory of the Susceptible-Infected-Recovered-Vaccinated
Epidemic Model (SIRV) for High Basic Reproduction Number}
\author{Oleg B. Shiryaev}
\email{shiryaev_ob@rsmu.ru}
\affiliation{Biomedicine Institute, Pirogov Russian National Research Medical University, Ostrovityanov Street 1, Moscow, Russia, 117513} 
\date{\today}

\begin{abstract}
The susceptible-infected-recovered-vaccinated model (SIRV) of epidemic
processes is treated for the combination of intercompartment transition
rates yielding a high basic reproduction number, a condition corresponding
to an intense epidemic. Multiple-timescale solutions to the SIRV model
equations are presented, with the inverse of the basic reproduction
number employed as the expansion parameter. The asymptotics emerge
in implicit form but translate into an explicit approximation for
the magnitude of the epidemic peak vs the vaccination rate. The findings
are shown to agree thoroughly with the numerical results for the SIRV
model. The latter becoming the susceptible-infected-recovered (SIR) one absent the vaccination, the asymptotic results  are reduced to those for the SIR model accordingly, with the expressions for compartment sizes acquiring explicit form.
\end{abstract}


\maketitle

\part*{Introduction }

Since the introduction of the susceptible-infected-recovered (SIR)
epidemic model \cite{OriginalSIRPaper}, compartmental models continue
to play an important role in simulating the dynamic of infectious
disease outbreaks (relatively recent examples may be found e.g. in
\cite{Tolles},\cite{Roda},\cite{ChongYouEstradaRef19}). The SIRV
model, which describes the dynamic exchanges between four presumed
compartments - the susceptible, the infected, the recovered, and the vaccinated
- is a SIR model upgrade, largely motivated by the vaccination campaign
implemented in the course of the past COVID epidemic \cite{Schlickeiser SIRV}.
In its framework, the recovered retain immunity indefinitely as in
the SIR one, and so do the vaccinated who are withdrawn from
the numbers of the susceptible at a given rate (it should be noted
that a broader model taking into account the loss of immunity among
the fraction of the vaccinated is also available \cite{Turkyilmazoglu SIRVI}).
Reader can find an integrative review on the subject of modeling in
epidemiology in \cite{HethcoteReview}. 

A class of multiple-timescale asymptotic solutions to the SIRV model
equations pertinent to the situation of high basic reproduction number,
which implies intense unfolding of the epidemic, is presented below. The resulting
asymptotic equations afford implicit
solutions, but an approximate expression in explicit form is derived from
them for the dependence of the maximal size of the infected compartment
over the epidemic on the vaccination rate for a given set of the initial
conditions. This expression may make it possible to assess the impact
of the vaccination campaign on the maximal workload to which the
healthcare system will be exposed. 

The SIRV model is easily converted into the SIR one assuming zero vaccination rate. Accordingly, the SIR asymptotic solutions and epidemic peak estimate  follow from those for the SIRV model in the corresponding limit and, moreover, the SIR solutions are found to have explicit form.    

\section{Model and Normalization}

The collision between a rising epidemic and the concurrent mass vaccination
is described by the following (SIRV) model

\begin{equation}
\frac{di}{dt}=\beta\,s\,i-\gamma\,i,\label{eq:SIRVdimensionalInfected}
\end{equation}

\begin{equation}
\frac{ds}{dt}=-\beta\,s\,i-\chi \,s,\label{eq:eq:SIRdimensionalSusceptible}
\end{equation}

\begin{equation}
\frac{dr}{dt}=\gamma\,i,\label{eq:eq:SIRdimensionalRecovered}
\end{equation}

\begin{equation}
\frac{dv}{dt}=\chi\,s.\label{eq:SIRVdimensionalVaccinated}
\end{equation}
Here $i$, $s$, $r$, and $v$ are the sizes of the infected, susceptible,
recovered, and vaccinated compartments of the population, with $\beta$,
$\gamma$, and $\chi$ standing for the infection, recovery, and vaccination
rates. Eqs. (\ref{eq:SIRVdimensionalInfected})--(\ref{eq:SIRVdimensionalVaccinated})
sustain the conservation law $i(t)+s(t)+r(t)+v(t)=N$, where $N$
is the total population size. 

The model may be rewritten in terms of the normalized time $\tau=\beta\,t$
and of the related functions $u(\tau)=i(t),\:\sigma(\tau)=s(t),\:\rho(\tau)=r(t)$,
$\phi(\tau)=v(t)$

\begin{equation}
u'(\tau)-\sigma(\tau)\,u(\tau)+\epsilon\,u(\tau)=0,\label{eq:SIRVdimensionlessInfected}
\end{equation}

\begin{equation}
\sigma'(\tau)+\sigma(\tau)\,u(\tau)+k\,\sigma(\tau)=0,\label{eq:SIRVdimensionlessSusceptible}
\end{equation}

\begin{equation}
\rho'(\tau)=\epsilon\,u(\tau),\label{eq:SIRVdimensionlessRecovered}
\end{equation}

\begin{equation}
\phi'(\tau)=k\,\sigma(\tau),\label{eq:SIRVdimensionlessVaccinated}
\end{equation}
where $k=\chi/\beta$ and $\epsilon=\gamma/\beta$. The latter quantity
equals the inverse of the basic reproduction number for the epidemic
and therefore makes a small parameter in case a sharp outbreak of
the disease is underway. Asymptotics of the solutions to Eqs. (\ref{eq:SIRVdimensionlessInfected})--(\ref{eq:SIRVdimensionlessRecovered})
in the small parameter $\epsilon$ are developed below.

Adding up Eqs. (\ref{eq:SIRVdimensionlessInfected}) and (\ref{eq:SIRVdimensionlessSusceptible})
yields the linear equation

\begin{equation}
\sigma'(\tau)+u'(\tau)+\epsilon\,u(\tau)+k\,\sigma(\tau)=0\label{eq:linearForSusceptibles}
\end{equation}
to replace Eq. (\ref{eq:SIRVdimensionlessInfected}). The set comprising
Eqs. (\ref{eq:SIRVdimensionlessSusceptible}) and  (\ref{eq:linearForSusceptibles}),
along with the dependent equations (\ref{eq:SIRVdimensionlessRecovered})
and (\ref{eq:SIRVdimensionlessVaccinated}), is utilized below as
an equivalent of the SIRV model embodied in Eqs. (\ref{eq:SIRVdimensionlessInfected})--(\ref{eq:SIRVdimensionlessVaccinated}). 

\section{Asymptotic Algorithm }

Asymptotic solutions to Eqs. (\ref{eq:SIRVdimensionlessInfected})--(\ref{eq:SIRVdimensionlessVaccinated})
can be developed as multiple-timescale series in $\epsilon$ of the form

\[
u(\tau)=u_{0}(\tau)+\varLambda u_{0}(\psi)+\epsilon\left(u_{1}(\tau)+\varLambda u_{1}(\psi)\right)+\ldots\,,
\]

\[
\sigma(\tau)=\sigma_{0}(\tau)+\varLambda\sigma_{0}(\psi)+\epsilon\left(\sigma_{1}(\tau)+\varLambda\sigma_{1}(\psi)\right)+\ldots\,,
\]

\[
\rho(\tau)=\rho_{0}(\tau)+\varLambda\rho_{0}(\psi)+\epsilon\left(\rho_{1}(\tau)+\varLambda\rho_{1}(\psi)\right)+\ldots\,,
\]

\[
\phi(\tau)=\phi_{0}(\tau)+\varLambda\phi_{0}(\psi)+\epsilon\left(\phi_{1}(\tau)+\varLambda\phi_{1}(\psi)\right)+\ldots\,,
\]
where the variables $\tau$ and $\psi=\epsilon\,\tau$ are the system's
fast and slow times. The above series being substituted into Eqs.
(\ref{eq:SIRVdimensionlessInfected})--(\ref{eq:SIRVdimensionlessVaccinated}),
those decompose into the following set of equations in terms of order
in $\epsilon$ and dependence on either $\tau$ or $\psi$

\begin{equation}
\left(k+\text{\ensuremath{\varLambda u_{0}}}(\psi)\right)\,\varLambda\sigma_{0}(\psi)=0,\label{eq:suscSlow0}
\end{equation}

\begin{equation}
\left(\varLambda\sigma_{0}(0)+\sigma_{0}(\tau)\right)u_{0}(\tau)+\left(k+\varLambda u_{0}(0)\right)\sigma_{0}(\tau)+\sigma_{0}'(\tau)=0,\label{eq:suscFast0}
\end{equation}

\begin{equation}
k\,\varLambda\sigma_{0}(\psi)=0,\label{eq:linSlow0}
\end{equation}

\begin{equation}
k\,\sigma_{0}(\tau)+u_{0}'(\tau)+\sigma_{0}'(\tau)=0,\label{eq:linearForSusceptiblesFast}
\end{equation}

\begin{equation}
\varLambda u_{1}(\psi)\,\varLambda\sigma_{0}(\psi)+\left(k+\varLambda u_{0}(\psi)\right)\varLambda\sigma_{1}(\psi)+\varLambda\sigma_{0}'(\psi)=0,\label{eq:suscSlow1}
\end{equation}

\begin{equation}
\varLambda u_{0}'(\psi)+\varLambda u_{0}(\psi)+\varLambda\sigma_{0}'(\psi)=0,\label{eq:linSlow1}
\end{equation}

\begin{equation}
\rho_{0}'(\tau)=0,\label{eq:recoveredFast0}
\end{equation}

\begin{equation}
\varLambda\rho_{0}'(\psi)=\varLambda u_{0}(\psi),\label{eq:recoveredSlow0}
\end{equation}

\begin{equation}
\phi_{0}'(\tau)=k\,\sigma_{0}(\tau).\label{eq:immunizedFast0}
\end{equation}
Eqs. (\ref{eq:suscSlow0}) and (\ref{eq:linSlow0}) show that
$\varLambda\sigma_{0}(\psi)=0$. It thereby follows from Eq. (\ref{eq:suscSlow1})
that $\varLambda\sigma_{1}(\psi)=0$. Consequently, Eq. (\ref{eq:linSlow1})
becomes

\begin{equation}
\varLambda u_{0}'(\psi)+\varLambda u_{0}(\psi)=0.\label{eq:infectedSlowDisaggregated}
\end{equation}
Eqs. (\ref{eq:suscFast0}) and (\ref{eq:linearForSusceptiblesFast})
combine into 

\begin{equation}
u_{0}'(\tau)=\left(u_{0}(\tau)+\varLambda u_{0}(0)\right)\sigma_{0}(\tau).\label{eq:susceptibleFastDisaggregated}
\end{equation}
Eqs. (\ref{eq:linearForSusceptiblesFast}), (\ref{eq:infectedSlowDisaggregated}),
and (\ref{eq:susceptibleFastDisaggregated}) comprise a set for computing
approximate solutions in the framework of model (\ref{eq:SIRVdimensionlessInfected})--(\ref{eq:SIRVdimensionlessVaccinated}).

\section{Implicit Solutions to Asymptotic Equations}

Eq. (\ref{eq:infectedSlowDisaggregated}) readily integrates to

\[
\varLambda u_{0}(\psi)=\varLambda u_{0}(0)\,e^{-\psi},
\]
while Eqs. (\ref{eq:linearForSusceptiblesFast}) and (\ref{eq:susceptibleFastDisaggregated})
invite solutions in implicit form given by

\[
u_{0}(\tau)=p(\theta),\:\sigma_{0}(\tau)=q(\theta),
\]

\[
\theta=\int_{0}^{\tau}\sigma_{0}(\mu)\,d\mu.
\]
The equations for the functions thus introduced 
are 

\[
\varLambda u_{0}(0)+p(\theta)=p'(\theta),\:k+p'(\theta)+q'(\theta)=0,
\]
where

\begin{equation}
\theta'(\tau)=i(0)\left(1-e^{\theta(\tau)}\right)-k\,\theta(\tau)+s(0),\label{eq:implicitVarEq}
\end{equation}
(subject to the condition 
$\theta (0) = 0$ for convenience) and the corresponding solutions make

\begin{equation}
p(\theta)=\left(p(0)+\varLambda u_{0}(0)\right)e^{\theta}-\varLambda u_{0}(0),\label{eq:pImplicit}
\end{equation}

\begin{equation}
q(\theta)=\varLambda u_{0}(0)-\left(p(0)+\varLambda u_{0}(0)\right)e^{\theta}-k\,\theta+p(0)+q(0).\label{eq:qImplicit}
\end{equation}
The initial conditions for the functions $p(\theta)$, $q(\theta)$,
and $\varLambda u_{0}(\psi)$ must be linked to those for $u(\tau)$
and $\sigma(\tau)$ and, thereby, for the SIRV model 
(\ref{eq:SIRVdimensionlessInfected})--(\ref{eq:SIRVdimensionlessVaccinated}). First, Eqs. (\ref{eq:pImplicit}) and (\ref{eq:qImplicit})
at $\tau=0$ yield 

\begin{equation}
\varLambda u_{0}(0)+p(0)=i(0),\label{eq:infectedInit}
\end{equation}

\begin{equation}
q(0)=s(0).\label{eq:susceptibleInit}
\end{equation}
Secondly, relevant conditions can be drawn from the requirement that
$u(\tau)$ and $\sigma(\tau)$ must vanish as the argument tends to
infinity. With the notation $\theta(\infty)=\theta_{\infty}$, the
summation of Eqs. (\ref{eq:pImplicit}) and (\ref{eq:qImplicit})
yields

\begin{equation}
p(0)+q(0)-\theta_{\infty}\,k=0.\label{eq:linkAtInfinity}
\end{equation}
Overall, the constants in Eqs. (\ref{eq:pImplicit}) and (\ref{eq:qImplicit})
are calculated by resolving Eqs. (\ref{eq:infectedInit})--(\ref{eq:linkAtInfinity})
along with the condition $p(\theta_{\infty})=0$. The result is

\[
\varLambda u_{0}(0)=k\,W\left(\frac{e^{\frac{i(0)+s(0)}{k}\,i(0)}}{k}\right),
\]
where $W$ is the Lambert function, with the other constants inferred
according to Eqs. (\ref{eq:infectedInit}) and (\ref{eq:susceptibleInit}).
Given the above, the ground-state solution to the problem embodied
in Eqs. (\ref{eq:SIRVdimensionlessInfected}) and (\ref{eq:SIRVdimensionlessSusceptible})
emerges in implicit form as 

\begin{equation}
u(\tau)=i(0)e^{\theta}+k\,W\left(\frac{e^{\frac{i(0)+s(0)}{k}\,i(0)}}{k}\right)\,(e^{-\psi}-1)+\ldots\,,\label{eq:infected}
\end{equation}

\begin{equation}
\sigma(\tau)=i(0)\left(1-e^{\theta}\right)-k\,\theta+s(0)+\ldots\,.\label{eq:susceptible}
\end{equation}
Based on Eqs. (\ref{eq:recoveredSlow0}) and (\ref{eq:immunizedFast0})
combined with Eqs. (\ref{eq:infected}) and (\ref{eq:susceptible}),
the lowest-order solutions for the recovered and immunized fractions are found
to be

\begin{equation}
\rho(\tau)=k\,W\left(\frac{e^{\frac{i(0)+s(0)}{k}\,i(0)}}{k}\right)\left(1-e^{-\psi}\right)+r(0)\ldots\,,\label{eq:recovered}
\end{equation}

\begin{equation}
\phi(\tau)=k\,\theta+\phi(0)\ldots\,.\label{eq:immunized}
\end{equation}
Eqs. (\ref{eq:infected})--(\ref{eq:immunized}) provide an implicit asymptotic solution to the SIRV model embodied
in Eqs. (\ref{eq:SIRVdimensionalInfected})--(\ref{eq:SIRVdimensionalVaccinated}),
with $\theta$ defined by Eq. (\ref{eq:implicitVarEq}).

\begin{figure}
  \centering
  \includegraphics[width=0.9\textwidth]{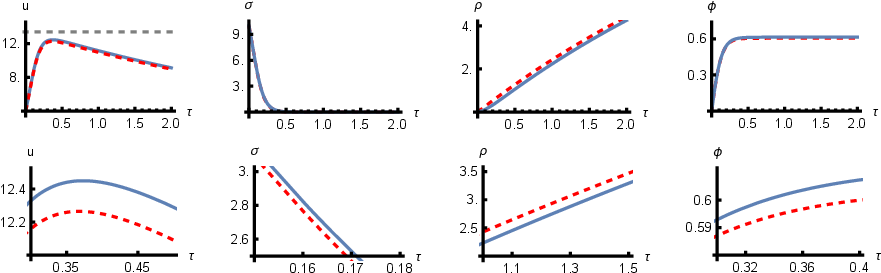}
  \caption{
  Numerical solutions to Eqs. (\ref{eq:SIRVdimensionlessInfected})--(\ref{eq:SIRVdimensionlessVaccinated}) (solid line)
and the corresponding asymptotic solution calculated according to
Eqs. (\ref{eq:infected})--(\ref{eq:immunized}) (dashed line). The
parameters in the depicted case are: $\epsilon=0.2$, $k=0.5$, $i(0)=4$,
$s(0)=10$, $r(0)=0$, $v(0)=0$. The grid line in the
plot for $u(\tau)$ marks the epidemic peak estimate given by Eq.
(\ref{eq:infectedPeakMagnitude}). The upper row shows plots for a
period of time long enough to reflect the epidemic peak and the plateau, and
the lower row - magnified views illustrating the sufficiency of precision
afforded by the asymptotic solution.
  }
  \label{ShiryaevOB_AsymptSIRV_Fig1}
\end{figure}

\begin{figure}
  \centering
  \includegraphics[width=0.9\textwidth]{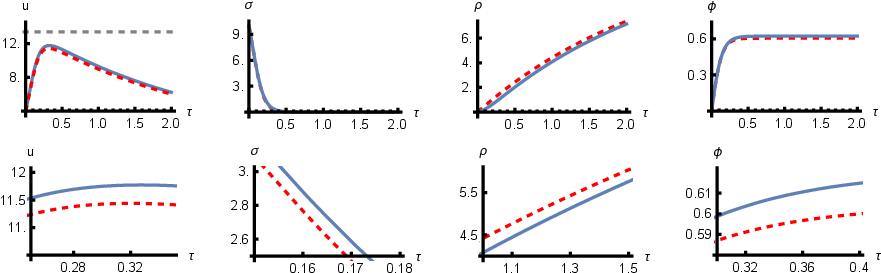}
  \caption{
  Numerical solutions to Eqs. (\ref{eq:SIRVdimensionlessInfected})--(\ref{eq:SIRVdimensionlessVaccinated})  (solid line)
and the corresponding asymptotic solution calculated according to
Eqs. (\ref{eq:infected})--(\ref{eq:immunized}) (dashed line). The
parameters in the depicted case are: $\epsilon=0.4$, $k=0.5$, $i(0)=4$,
$s(0)=10$, $r(0)=0$, $v(0)=0$. The grid line in the
plot for $u(\tau)$ marks the epidemic peak estimate given by Eq.
(\ref{eq:infectedPeakMagnitude}). The upper row shows plots for a
period of time long enough to reflect the epidemic peak and the plateau, and
the lower row - magnified views illustrating the sufficiency of precision
afforded by the asymptotic solution. 
  }
  \label{ShiryaevOB_AsymptSIRV_Fig2}
\end{figure}

Asymptotic solutions (\ref{eq:infected})-(\ref{eq:immunized}) and
numerical solutions to Eqs. (\ref{eq:SIRVdimensionlessInfected})--(\ref{eq:SIRVdimensionlessVaccinated})
are juxtaposed by overlaying the corresponding curves in Figs. \ref{ShiryaevOB_AsymptSIRV_Fig1}
and 
\ref{ShiryaevOB_AsymptSIRV_Fig2}. The parameters of the calculations are listed in the figure
captions. The first row in each figure, spanning the sharp rise of
the epidemic process and the subsequent protracted plateau, demonstrates
that the asymptotic solutions suggested above practically coincide
with the numerical results. The second row consists of magnified excerpts
revealing the minute difference between the numerical and asymptotic
curves, which is imperceptible at the scale adopted throughout the
first row.

\section{Dependence of the Epidemic Peak Magnitude on the Vaccination Rate}

The dependence of the epidemic peak magnitude on the vaccination rate
$k$ can be inferred approximately in the framework of the SIRV model
(\ref{eq:SIRVdimensionlessInfected})--(\ref{eq:SIRVdimensionlessSusceptible})
with the help of the asymptotic solution (\ref{eq:infected})--(\ref{eq:immunized}).
It follows from Eqs. (\ref{eq:infected}) and (\ref{eq:implicitVarEq})
to the lowest order in $\epsilon$ that the infected compartment reaches
its maximal size for 

\[
\theta_{0}=\frac{i(0)+s(0)-k\,W\left(\frac{e^{\frac{i(0)+s(0)}{k}}\,i(0)}{k}\right)}{k}
\]
(it should be noted that the same equation is obtained under pertinent
approximations from the condition $\sigma_{0}(\tau)=0$). With the
above substituted into Eq. (\ref{eq:infected}), the maximal size
of the infected compartment, similarly to the lowest order in $\epsilon$, is
found to be

\begin{equation}
i_{\max}(k)=k\,W\left(\frac{e^{\frac{i(0)+s(0)}{k}}\,i(0)}{k}\right).\label{eq:infectedPeakMagnitude}
\end{equation}
Given Eq. (\ref{eq:infectedPeakMagnitude}), the above asymptotic expressions
may be modified by reducing Eqs. (\ref{eq:infected}) and (\ref{eq:recovered})
to the compact form

\begin{equation}
u(\tau)=i(0)e^{\theta}-i_{\max}(k)\left(1-e^{-\psi}\right)+\ldots\,,
\label{eq:infectedCompact}
\end{equation}

\begin{equation}
\rho(\tau)=i_{\max}(k)\left(1-e^{-\psi}\right)+r(0)\ldots\,.
\label{eq:recoveredCompact}
\end{equation}

\begin{figure}
  \centering
  \includegraphics[width=0.5\textwidth]{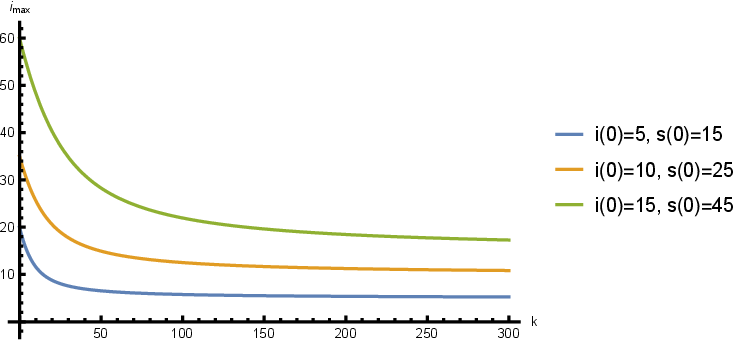}
  \caption{
  Epidemic peak magnitudes vs the normalized vaccination rate
$k$ according to Eq. (\ref{eq:infectedPeakMagnitude}) for various initial conditions of problem (\ref{eq:SIRVdimensionlessInfected})--(\ref{eq:SIRVdimensionlessVaccinated}).
  }
  \label{ShiryaevOB_AsymptSIRV_Fig3}
\end{figure}

\begin{figure}
  \centering
  \includegraphics[width=0.9\textwidth]{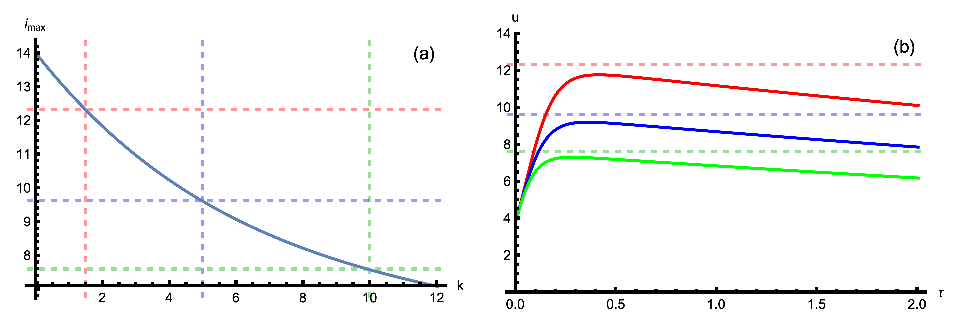}
  \caption{
  Epidemic peak magnitude estimated using Eq. (\ref{eq:infectedPeakMagnitude})
vs the normalized vaccination rate $k$ for $\epsilon=0.1$, $i(0)=4$,
$s(0)=10$, $r(0)=0$, $v(0)=0$ (a) and the calculated
sizes of the infected compartment for these parameters and $k=1.5,\,5,\,10$,
with the peak estimates  shown as dashed grid lines (b).
  }
  \label{ShiryaevOB_AsymptSIRV_Fig4}
\end{figure}

For a specific set of initial conditions, attainable peak values for
the infected compartment are confined to the interval

\[
i(0)=i_{\max}(\infty)<u_{\max}(k)\leq i_{\max}(0),
\]
where 
\begin{equation}
i_{\max}(0)=i(0)+s(0).
\label{eq:SIRPeak}
\end{equation}
The above maximal size of the infected compartment for $k = 0$  occurs naturally in the SIR limit. The dependence $i_{\max}(k)$
is exemplified for various initial conditions in 
Fig.~\ref{ShiryaevOB_AsymptSIRV_Fig3}. 
Eq. (\ref{eq:infectedPeakMagnitude})
provides an assessment of the vaccination campaign efficiency in partially
suppressing the epidemic peak. Fig.~\ref{ShiryaevOB_AsymptSIRV_Fig4} illustrates the impact of
the vaccination rate on the magnitude of the epidemic peak for a particular
set of the initial conditions and for several values of $k$. The
figure also highlights the agreement between the approximate estimates
based on Eq. (\ref{eq:infectedPeakMagnitude}) and the specific solutions
to Eqs. (\ref{eq:infected})--(\ref{eq:immunized}), which holds
within the precision of the asymptotic method. 

Due to the zeroth-order expansion in the small parameter $\epsilon$,
the estimate stemming from Eq. (\ref{eq:infectedPeakMagnitude}) emerges
as independent of the latter, but it remains workable as long as the
parameter $\epsilon$ is small. The gridlines in the $u(\tau)$ plots
in Figs.~\ref{ShiryaevOB_AsymptSIRV_Fig1} 
and~\ref{ShiryaevOB_AsymptSIRV_Fig2} correspond to the peak estimates projected
based on Eq. (\ref{eq:infectedPeakMagnitude}), and apparently they
stay adequate in both cases even though the values of $\epsilon$
in the cases considered are different. 

\section{Reduction to SIR Model}

SIRV model (\ref{eq:SIRVdimensionlessInfected})--(\ref{eq:SIRVdimensionlessVaccinated}) being reduced to the SIR one by setting 
$k=0$, the above results apply to the latter entirely under the same  assumption. Eq. (\ref{eq:implicitVarEq}) integrates easily in the case so that
\[
\theta(\tau)=\log \left(\frac{(i(0)+s(0)) e^{  (i(0)+s(0))\tau}}{i(0)
   e^{  (i(0)+s(0))\tau}+s(0)}
   \right)\] 
   and the substitution of the expression, along with Eq. (\ref{eq:SIRPeak}), into Eqs. (\ref{eq:infectedCompact}), (\ref{eq:susceptible}), and (\ref{eq:recoveredCompact}) leads to the SIR asymptotic solutions, which, in contrast to the general SIRV ones,  are explicit
   
       \begin{align}
  u(\tau)
&=
-\frac{s(0)\left(i(0)+s(0)\right)}{s(0)+i(0)\exp\left(\left(i(0)+s(0)\right)\tau\right)} + 
\left(i(0)+s(0)\right)\exp(-\psi) + \ldots \,
,\label{eq:asymptInfected} \\
 \sigma(\tau)
&=
\frac{s(0)\left(i(0)+s(0)\right)}{s(0)+i(0)\exp\left(\left(i(0)+s(0)\right)\tau\right)} + \ldots \,
,\label{eq:asympSusceptibles} \\
 r(\tau)
&=
N-\left(i(0)+s(0)\right)\exp(-\psi) + \ldots \,
.\label{eq:asymptRecovered}
\end{align}

\section*{Conclusions}

The four-compartment susceptible-infected-recovered-vaccinated (SIRV)
epidemic model is treated in the present study. The assumptions underlying
the model are that both recovered and vaccinated remain steadily immune
while the compartments interact and evolve as the disease spreads. Asymptotic
solutions to the equations of the SIRV model are developed for the
case where the infection rate substantially exceeds the recovery rate.
The disparity between the constants implies a high basic reproduction
number, placing the system in the epidemic outbreak regime. Multiple-timescale
solutions to the SIRV model equations are obtained, with the inverse
of the basic reproduction number playing the role of the expansion
parameter. The solutions which describe the sizes of the four compartments
vs time are derived in implicit form as all of them are expressed
via a function which solves an auxiliary differential equation free
of the small parameter of the problem. Nevertheless, an approximate
relation between the size of the infected compartment at the time
of the epidemic peak and the vaccination rate is established on the
basis of the above. This relation may serve to predict to what degree
the maximal workload on the healthcare system over the epidemic would
be reduced for a given vaccination campaign intensity. 

The asymptotic
results are practically coincidental with those obtained by solving
the SIRV equations numerically.  Explicit asymptotic expressions for the compartment sizes vs. time and the corresponding epidemic peak estimate are established for the particular case without vaccination, which corresponds to the SIR model.


\begin{thebibliography}{1}

\bibitem{OriginalSIRPaper}
W.O.~Kermack and A.G.~McKendrick. 
A contribution to the mathematical theory of epidemics.
{\em Bulletin of Mathematical Biology. Proceedings of the Royal Society of London. Series A, Containing Papers of a Mathematical and Physical Character,}
115(772):700-721, 1927.

\bibitem{Tolles} 
Tolles J, Luong T. Modeling Epidemics With Compartmental
Models. {\em JAMA}.   323(24):2515-2516, 2020 doi: 10.1001/jama.2020.8420.
PMID: 32459319. 

\bibitem{Roda}
W.C.~Roda, M.B.~Varughese, D.~Han, M.Y.~Li. 
Why is it difficult to accurately predict the COVID-19 epidemic? 
{\em Infect Dis Model., } 
5:271-281, 2020.
PMID: 32289100; PMCID: PMC7104073.

\bibitem{ChongYouEstradaRef19} 
C.~You, Y.~Deng, W.~Hu, J.~Sun, Q.~Lin, F.~Zhou, 
Cheng~Heng~Pang, Yuan~Zhang, Zhengchao~Chen, 
Xiao-Hua~Zhou.
Estimation of the time-varying reproduction number of COVID-19 outbreak in China. 
{\em Int. J. Hyg. Environ. Health,} 
 228:113555, 2020. 
May 11. PMID:
32460229; PMCID: PMC7211652.

\bibitem{Schlickeiser SIRV} 
Schlickeiser, R.; Krцger, M. Analytical
Modeling of the Temporal Evolution of Epidemics Outbreaks Accounting
for Vaccinations. {\em Physics}  3, 386-426, 2021. https://doi.org/10.3390/physics3020028.

\bibitem{Turkyilmazoglu SIRVI}
Mustafa Turkyilmazoglu. An extended
epidemic model with vaccination: Weak-immune SIRVI, {\em Physica A: Statistical
Mechanics and its Applications}, Volume 598,  127429, 2022 
ISSN 0378-4371,
https://doi.org/10.1016/j.physa.2022.127429.

\bibitem{HethcoteReview}
H.W.~Hethcote. 
The mathematics of infectious diseases.
{\em SIAM Review,} 
42(4):599--653, 2000. 


\end{thebibliography}
\end{document}